\documentclass[english]{article}
\usepackage{jcappub}
\usepackage{lmodern}

\usepackage{afterpage}
\usepackage{bm}        

\usepackage{slantsc}
\usepackage{graphicx}
\DeclareGraphicsExtensions{.pdf,.png,.jpg}

\def\d{\mathrm d}

\def\Om{\hat{\mathbf{u}}}
\def\OGW{\Omega_\text{GW}}
\def\OmComp{\Omega_\alpha}
\def\orf{\gamma_{\text{iso}}}

\begin{document}

\title{Component Separation of a Isotropic Gravitational Wave Background}

\author[a]{Abhishek Parida}
\author[b]{Sanjit Mitra}{
\author[a]{and Sanjay Jhingan}

\emailAdd{abhishekparida22@gmail.com}
\emailAdd{sanjit@iucaa.ernet.in}
\emailAdd{sjhingan@jmi.ac.in}

\affiliation[a]{Centre for Theoretical Physics, Jamia Millia Islamia, New Delhi 110025, India}
\affiliation[b]{IUCAA, P. O. Bag 4, Ganeshkhind, Pune 411007, India}

\keywords{Gravitational Waves}

\abstract{
A Gravitational Wave Background (GWB) is expected in the universe from the superposition of a large number of unresolved astrophysical sources and phenomena in the early universe. Each component of the background (e.g., from primordial metric perturbations, binary neutron stars, milli-second pulsars etc.) has its own spectral shape. Many ongoing experiments aim to probe GWB at a variety of frequency bands. In the last two decades, using data from ground-based laser interferometric gravitational wave~(GW) observatories, upper limits on GWB were placed in the frequency range of $\sim 50-1000$~Hz, considering one spectral shape at a time. However,
one strong component can significantly enhance the estimated strength of another component. Hence, estimation of the amplitudes of the components with different spectral shapes should be done jointly. Here we propose a method for ``component separation'' of a statistically isotropic background, that can, for the first time, jointly estimate the amplitudes of many components and place upper limits. The method is rather straightforward and needs negligible amount of computation. It utilises the linear relationship between the measurements and the amplitudes of the actual components, alleviating the need for a sampling based method, e.g., Markov Chain Monte Carlo (MCMC) or matched filtering, which are computationally intensive and cumbersome in a multi-dimensional parameter space. Using this formalism we could also study how many independent components can be separated using a given dataset from a network of current and upcoming ground based interferometric detectors.
}

\maketitle

\section{Introduction}

Gravitational waves (GW) is one of the most exciting predictions of Einstein's theory of General Relativity (GR)~\cite{gw,MTW,ThorneG300}. Though there is convincing evidence of the existence of GW~\cite{HulseTaylor,WeisbergNiceTaylor}, direct detection of GW via man-made instruments has not been possible yet. Several experiments are being developed and proposed to detect GW~\cite{AdvLIGO,AdvVirgo,KAGRA,BICEP2,PlanckPolDust, BICEP2-Planck,IPTA,eLISA, DECIGO, BBO}, which will not only be an important test of GR by itself, but it promises a whole new set of observational windows to the universe. The ground based laser interferometric detectors will likely be the first ones to detect GW from individual events and the most promising sources for the first detection are the compact binary coalescences~\cite{LIGOEventRate}. GW astronomy, however, promises to probe many different kinds of sources, gravitational wave background (GWB) is one of the most interesting ones.
GWB can be created by the superposition of a large number of unresolved astrophysical sources~\cite{CowTan06,TanVuk08} (e.g., inspiralling compact binary stars, supernovae, millisecond pulsars, magnetars etc.) and phenomena in the early universe~\cite{AllenSchool,Grishchuk00,Turner96} (e.g., phase transition, inflation etc.). If the dominant sources are at high redshifts, the background is expected to be statistically isotropic~\cite{Mazumder2014}. A large number of experiments, present and future, aim to probe GWB at a variety of frequency bands. The advanced ground based laser interferometric observatories are expected to detect or tightly constrain a  GWB~\cite{TanVuk08,sgwbmdc15}.

Each astrophysical and cosmological source of GWB can create a background with a specific spectral shape~\cite{CowTan06}, as they are generated by different physical processes or distributions. Therefore, in general, the observed GWB is a superposition of different components. Since the backgrounds are generated by incoherent set of sources or events, the superposition is linear in energy density. A number of observational limits have been placed on GWB for specific spectral indices~\cite{BBNBound,SmithEtAl,RottiSouradeep,WMAP9par,PlanckParam,sgwbS5iso,isoLVC_2009-10,sgwbS5dir,EPTA_Limit}, but all considering only one spectral index at a time (even when the spectral index is considered to be a free parameter,~\cite{sgwbS5iso,sgwbmdc15} the search still assumes a single index background). However, the measured energy density of one component can be strongly influenced by the others, as they are derived from the same set of experimental data. Even more so if the searches were performed in overlapping frequency bands. Hence, estimation of GWB energy density must be done jointly for all the components. This is the main motivation of this paper.

Ours is not the first work to suggest this. Methods have been proposed and implemented to jointly constrain parameters characterising {\em two component backgrounds}~\cite{UngarelliVecchio03,MandicEtAl12}. However, these methods relied upon sampling of the parameter space, posing a challenge to extend those implementations beyond two components, as it would require dealing with a Likelihood function in a multi-dimensional parameter space. Here we propose a novel method that, utilising the linear relationship between the observed data and the amplitudes of the components, transforms the Maximum Likelihood parameter estimation problem to a matrix inversion problem, where the dimension of the matrix is rather small---equal to the number of components. This makes estimation of amplitudes of many components trivial, as long as the data contains enough information. We also investigate how many components can be estimated simultaneously from a given set of data.

This paper is organised as follows. Our method of component separation is described in section~\ref{sec:method}. We demonstrate the method for two toy cases using realistic parameters in section~\ref{sec:toy}. An investigation to address how many components can be separated from a given dataset is presented in section~\ref{sec:indcomp}. We conclude with discussions in section~\ref{sec:concl}.

\section{Formalism}
\label{sec:method}

\subsection{General characteristics of signal from a background}

First we summarise certain primary results for isotropic backgrounds from standard literature~\cite{Michelson87,christ92,flan93,allen01}. A stochastic background is created by incoherent superposition of GW from several sources. If $h_1(t)$ and $h_2(t)$ are the time series strain signals created in two detectors ($1$ and $2$) and $\widetilde{h}_1(t;f)$, $\widetilde{h}_2(t;f)$ are their respective (discrete) short-term Fourier transforms (SFTs) over time segments of duration $\Delta T$, where, in the SFTs, $t$ represents a time-stamp that uniquely marks a segment, one can write
\begin{equation}
\langle \widetilde{h}^*_1(t;f) \, \widetilde{h}_2(t';f') \rangle \ = \ \Delta T \delta_{tt'} \delta_{ff'} \, H(f) \, \orf(f) \, .
\label{eq:h1h2base}
\end{equation}
Here $\orf(f)$ is the overlap reduction function for a statistically isotropic background defined in terms of the detector antenna pattern functions $F_{+,\times}^{1,2}(t,\Om)$ as,
\begin{equation}
\orf(f) \ = \ \frac{5}{8\pi} \int_{4\pi} \d\Om \, [F_{+}^{1}(t,\Om) F_{+}^{2}(t,\Om) + F_{\times}^{1}(t,\Om) F_{\times}^{2}(t,\Om)] \, e^{2 \pi i \mathbf{\Delta x}(t) \cdot \Om/c} \, ,
\end{equation}
where $\Om$ is a direction on the sky and $\mathbf{\Delta x}(t)$ is the separation vector between the two detector sites in the Celestial Equatorial frame whose origin coincides with the center of the Earth. $H(f)$ is the two-sided power spectral density (PSD) of the background, which is related to the more popular quantity used for characterising a stochastic background, $\OGW(f)$, through
\begin{equation}
H(f) \ = \ \frac{3 H_0^2}{32 \pi^3} \, |f|^{-3} \, \OGW(|f|) \, .
\label{eq:HOmega}
\end{equation}
$\OGW(|f|)$ is defined as the energy density ($\rho_\text{GW}$) per unit logarithmic frequency interval,
\begin{equation}
\OGW(f) \ = \ \frac{1}{\rho_\text{c}} \frac{\d \rho_\text{GW}}{\d \ln f} \, ,
\end{equation}
expressed in the units of critical energy density needed for a spatially flat universe,
\begin{equation}
\rho_\text{c} \ = \ \frac{3 H_0^2 c^2}{8 \pi G} \, ,
\end{equation}
where $H_0$ is the Hubble constant.

In addition to their respective strain signal, each detector also has nearly uncorrelated random noise $n_1(t)$ and $n_2(t)$, with corresponding SFTs $\widetilde{n}_1(t;f)$ and $\widetilde{n}_2(t;f)$. SFTs of the total observed signals $s_1(t)$ and $s_2(t)$ in the detectors are simply
\begin{eqnarray}
\widetilde{s}_1(t;f) &=& \widetilde{h}_1(t;f) \ + \ \widetilde{n}_1(t;f) \, , \\
\widetilde{s}_2(t;f) &=& \widetilde{h}_2(t;f) \ + \ \widetilde{n}_2(t;f) \, .
\end{eqnarray}
In the small signal limit, where the variance of the strain signal is much less than the variance of the noise, one can write
\begin{equation}
\widetilde{s}^*_1(t;f) \, \widetilde{s}_2(t';f') \ \approx \ \delta_{tt'} \delta_{ff'} \Delta T \, \frac{3 H_0^2}{20 \pi^2}  \, \orf(f) \, |f|^{-3} \, \OGW(|f|) \ + \ \widetilde{n}^*_1(t;f) \, \widetilde{n}_2(t';f') \, .
\label{eq:s1s2}
\end{equation}

\subsection{A background with multiple components}
\label{subsec:derivation}

If the background was created by multiple components, and the component $\alpha$ has a {\em known} spectral shape $\mathcal{F}^\alpha(f)$ (e.g., a power-law, $\mathcal{F}^\alpha(f) = (f/f_\text{ref})^\alpha$, where $f_\text{ref}$ is a reference frequency, say, $100$Hz) with an amplitude of $\Omega_\alpha$, that is,
\begin{equation}
\OGW(f) \ = \ \sum_\alpha \Omega_\alpha \, \mathcal{F}^\alpha(f) \, ,
\end{equation}
one could rewrite Eq.~(\ref{eq:s1s2}) for $t=t'$ and $f=f'$ as a linear convolution equation,
\begin{equation}
\mathbf{C} \ = \ \mathbf{K} \cdot \mathbf{\Omega} \ + \ \mathbf{N}  \qquad \text{or}, \ C_{tf} \ = \ \sum_\alpha K^\alpha_{tf} \, \OmComp \ + \ N_{tf} \, ,
\label{eq:conv}
\end{equation}
where,
\begin{eqnarray}
\mathbf{\Omega} & \equiv & \Omega_\alpha \, , \\
\mathbf{C} & \equiv & C_{tf} \ := \ \widetilde{s}^*_1(t;f) \, \widetilde{s}_2(t;f) \, ,\\
\mathbf{N} & \equiv & N_{tf} \ := \ \widetilde{n}^*_1(t;f) \, \widetilde{n}_2(t;f) \, ,\\
\mathbf{K} & \equiv & K^\alpha_{tf} \ := \  \Delta T \, \left( \frac{3 H_0^2}{20 \pi^2} \right) \orf(f) \, |f|^{-3} \, \mathcal{F}^\alpha(|f|) \, .
\end{eqnarray}
If the one-sided noise PSD of the data segments from the two detectors are denoted by $P_1(t;f)$ and $P_2(t;f)$ respectively, the covariance of $N_{tf}$ becomes
\begin{equation}
\bm{\mathcal{N}} \ \equiv \ \mathcal{N}_{tft'f'} \ = \ \langle N_{tf}^* \, N_{t'f'} \rangle \ = \ \delta_{tt'} \delta_{ff'} \frac{(\Delta T)^2}{4} P_1(t;|f|) P_2(t;|f|) \, .
\label{eq:noiseCov}
\end{equation}
Since noise in the detectors $n_{1,2}(t)$ are nearly independent, $N_{tf}$ is of zero mean. Moreover, it is reasonable to assume that $N_{tf}$ follows a Gaussian distribution. This is because the analysis is usually done by collecting data over few hundreds of days. Since the ``kernel'' $\mathbf{K}$ in a GWB search  repeats after a time period of one sidereal day, noise is effectively averaged over an ensemble of few hundred elements~\cite{folding15}. Thus, one can impose the central limit theorem to argue that $N_{tf}$ follows a Gaussian distribution. In case of the search for a isotropic background, $\mathbf{K}$ is independent of time, hence the averaging can be done over all the segments together, taking the distribution even closer to a Gaussian.

\subsection{Joint estimation of amplitudes of multiple components}
\label{sec:jointEst}

Our main aim is to estimate $\mathbf{\Omega}$ from observed data.  The convolution equation, Eq.~(\ref{eq:conv}),  has a standard Maximum Likelihood solution for $\mathbf{\Omega}$, given by
\begin{equation}
\hat{\mathbf{\Omega}} \ = \ \mathbf{\Gamma}^{-1} \cdot \mathbf{X} \, ,
\label{eq:deconv}
\end{equation}
where,
\begin{eqnarray}
\mathbf{X} & = & \mathbf{K}^\dagger \cdot \bm{\mathcal{N}}^{-1} \cdot \mathbf{C} \, \label{eq:X} \\
\mathbf{\Gamma} & = & \mathbf{K}^\dagger \cdot \bm{\mathcal{N}}^{-1} \cdot \mathbf{K} \, . \label{eq:Gamma}
\end{eqnarray}
Note that, $\hat{\mathbf{\Omega}}$ is an unbiased estimator of the amplitudes $\mathbf{\Omega}$.
$\mathbf{\Gamma}$ is the Fisher information matrix for the estimated values and its inverse is the noise covariance matrix, $\mathbf{\Sigma} := \mathbf{\Gamma}^{-1}$.

Since the noise covariance matrix of the SFTs, $\bm{\mathcal{N}}$, is diagonal in time and frequency [Eq.~(\ref{eq:noiseCov})], one can write
\begin{equation}
[\bm{\mathcal{N}}^{-1}]_{tft'f'} \ = \ \delta_{tt'} \delta_{ff'} \frac{4}{(\Delta T)^2} \frac{1}{P_1(t;|f|) P_2(t;|f|)} \, .
\end{equation}
Substituting this in Eqs.~(\ref{eq:X} \& \ref{eq:Gamma}) one gets the final expressions to compute all $\hat{\Omega}_\alpha$,
\begin{eqnarray}
X_\alpha &=& \frac{4}{\Delta T} \left( \frac{3 H_0^2}{20 \pi^2} \right) \sum_{tf} \orf(|f|) \, \frac{\mathcal{F}^\alpha(|f|) \, \widetilde{s}^*_1(t;f) \, \widetilde{s}_2(t;f)}{|f|^3 \, P_1(t;|f|) P_2(t;|f|)} \, ,
\label{eq:unnormX}\\
\Gamma_{\alpha\beta} &=&  4 \, \left( \frac{3 H_0^2}{20 \pi^2} \right)^2 \sum_{tf} |\orf(|f|)|^2 \, \frac{\mathcal{F}^\alpha(|f|) \, \mathcal{F}^\beta(|f|)}{f^6 \, P_1(t;|f|) P_2(t;|f|)} \, .
\label{eq:unnormGamma}
\end{eqnarray}
The summation on the right of Eq.(\ref{eq:unnormGamma}) appears several times in the text, so from now on we will use a shorthand,
\begin{equation}
M_{\alpha\beta} \ := \ \sum_{tf} |\orf(|f|)|^2 \, \frac{\mathcal{F}^\alpha(|f|) \, \mathcal{F}^\beta(|f|)}{f^6 \, P_1(t;|f|) P_2(t;|f|)} \, .
\end{equation}

\subsection{Relation to the standard single component analysis}

Above algebra provides a complete framework needed for joint estimation of the components starting from raw data. However, for added convenience, we rescale the quantities $X_\alpha$ and $\Gamma_{\alpha\beta}$ to relate the joint analysis to the standard single component analysis.

If only one component $\alpha$ was present in the universe, substituting Eqs.~(\ref{eq:h1h2base} \& \ref{eq:HOmega}) in the expectation of Eq.~(\ref{eq:unnormX}), one gets,
\begin{equation}
\langle X_\alpha \rangle \ = \  4 \, \left( \frac{3 H_0^2}{20 \pi^2} \right)^2 \, M_{\alpha\alpha} \, \OmComp \ = \  \Gamma_{\alpha\alpha} \, \OmComp \, .
\end{equation}
Thus, the point estimate
\begin{equation}
\mathbf{Y} \ \equiv \ Y_\alpha \ := \ \Gamma_{\alpha\alpha}^{-1} \, X_\alpha
\end{equation}
is an unbiased estimator of $\OmComp$, that is, $\langle Y_\alpha \rangle = \OmComp$, {\em when only one component is present}, and is identical to the point estimate used in the standard single component search for an isotropic background~\cite{allen01,isoLVC_2009-10}. In general, however, $Y_\alpha$ is not an unbiased estimator of $\OmComp$ (though $\hat{\Omega}_\alpha$ is). Putting all together, we arrive at a key relation, 
\begin{equation}
\langle Y_\alpha \rangle \ = \ \sum_\beta B_{\alpha\beta} \, \Omega_\beta \, ,
\label{eq:normconv}
\end{equation}
where,
\begin{equation}
B_{\alpha\beta} \ := \ \frac{\Gamma_{\alpha\beta}}{\Gamma_{\alpha\alpha}} \ = \ \frac{M_{\alpha\beta}}{M_{\alpha\alpha}} \ = \ \frac{\sum_{tf} \frac{|\orf(|f|)|^2 \, \mathcal{F}^\alpha(|f|) \, \mathcal{F}^\beta(|f|)}{f^6 \, P_1(t;|f|) P_2(t;|f|)}}{\sum_{tf} \frac{|\orf(|f|)|^2 \,  \mathcal{F}^\alpha(|f|) \, \mathcal{F}^\alpha(|f|)}{f^6 \, P_1(t;|f|) P_2(t;|f|)}}  \, .
\label{eq:defB}
\end{equation}
Substituting these in the general (multi-component) solution, Eqs.~(\ref{eq:unnormX}, \ref{eq:unnormGamma} \& \ref{eq:deconv}), the final set of formulae needed for single and joint estimation of components from real data become,
\begin{eqnarray}
Y_\alpha &=& \left(\frac{20 \pi^2}{3 H_0^2} \right) \frac{1}{\Delta T \, M_{\alpha\alpha}} \sum_{tf} \orf(|f|) \, \frac{\mathcal{F}^\alpha(|f|) \, \widetilde{s}^*_1(t;f) \, \widetilde{s}_2(t;f)}{|f|^3 \, P_1(t;|f|) P_2(t;|f|)} \, ,
\label{eq:normX}\\
\hat{\Omega}_\alpha &=& \sum_\beta [B^{-1}]_{\alpha\beta} \, Y_\beta \, , \label{eq:normdeconv} \\
\Sigma_{\alpha\beta} &=& \frac{1}{4} \, \left( \frac{20 \pi^2}{3 H_0^2} \right)^2 [M^{-1}]_{\alpha\beta} \, .
\label{eq:normGamma}
\end{eqnarray}
Variance of the single component estimator, $\sigma^2_\alpha$, is related to the variance of the joint estimator (the diagonal elements of the covariance matrix), $\Sigma_{\alpha\alpha}$, through
\begin{equation}
\sigma^2_{\alpha} = \frac{1}{4} \left( \frac{20 \pi^2}{3 H_0^2} \right)^2 \frac{1}{M_{\alpha\alpha}} \ = \ \frac{1}{[M^{-1}]_{\alpha\alpha} M_{\alpha\alpha}} \, \Sigma_{\alpha\alpha} \, .
\end{equation}
Not surprisingly, they are identical for a single component search, as in that case $[M^{-1}]_{\alpha\alpha} = 1/M_{\alpha\alpha}$.

\subsection{Numerical implementation}

In practice, however, inversion of the matrix $\mathbf{B}$ introduces large numerical errors in the estimates $\hat{\mathbf{\Omega}}$. This is because even though the diagonal components of $\mathbf{B}$ are all unity, the off diagonal components can take any (small/large/positive/negative) values. We resolve this in the numerical implementation by making one more transformation to ``precondition'' the matrix. We define,
\begin{eqnarray}
\mathbf{Y}' \ \equiv \ Y'_\alpha &=& Y_\alpha \sqrt{M_{\alpha\alpha}} \, , \\
\hat{\mathbf{\Omega}}' \ \equiv \ \hat{\Omega}'_\alpha &=& \hat{\Omega}_\alpha \sqrt{M_{\alpha\alpha}} \, .
\end{eqnarray}
Then Eq.(\ref{eq:normdeconv}) can be transformed to,
\begin{equation}
Y'_\alpha \ = \ \sqrt{M_{\alpha\alpha}} \sum_\beta B_{\alpha\beta} \, \hat{\Omega}_\beta \ = \ \sum_\beta \frac{M_{\alpha\beta}}{\sqrt{M_{\alpha\alpha}} \sqrt{M_{\beta\beta}}} \, \hat{\Omega}'_\beta \ = \ \sum_\beta B'_{\alpha\beta} \, \hat{\Omega}'_\beta \, , \label{eq:primedConv}
\end{equation}
where,
\begin{equation}
\mathbf{B}' \ \equiv \ B'_{\alpha\beta} \ := \ \frac{M_{\alpha\beta}}{\sqrt{M_{\alpha\alpha}} \sqrt{M_{\beta\beta}}} \ = \ \frac{\sqrt{M_{\alpha\alpha}}}{\sqrt{M_{\beta\beta}}} \, B_{\alpha\beta} \, .
\end{equation}
Thus the redefined coupling matrix $B'_{\alpha\beta}$ has essentially become a ``normalised scalar product'' for different component pairs, which has a unit diagonal and the off-diagonal components are smaller positive numbers. Preconditioning of the matrix can reduce large numerical errors when it is inverted. Therefore, in the numerical implementation, we first solve Eq.~(\ref{eq:primedConv}) for $\hat{\Omega}'_\alpha$ by inverting the matrix $\mathbf{B}'$ and then convert it to $\hat{\Omega}_\alpha = \hat{\Omega}'_\alpha/\sqrt{M_{\alpha\alpha}}$.

Note that, while without preconditioning it may not be possible to invert a matrix numerically, it does not assure either that the convolution equation will become invertible, as two rows or columns of the $B'_{\alpha\beta}$ matrix could still be very close, leading to a nearly vanishing determinant. Invertibility of the problem depends on the characteristics of the detectors and the spectral shapes of the components that are being probed, which is discussed in more details in section~\ref{sec:indcomp}.

To apply our method on real data one can start from the point estimates ($Y_\alpha$) of the single index analysis (run separately, of course) and jointly estimate $\OmComp$ using Eq.~(\ref{eq:normdeconv}) going through the preconditioning process. Essentially one needs to invert the matrix $\mathbf{B}'$, whose dimension is equal to the number of components. So, in practice, $\mathbf{B}'$ is a few by few matrix, inverting which is computationally trivial, as long as it is invertible. One would however need to compute $\mathbf{B}'$ and $\mathbf{\Sigma}$, which only requires the noise PSDs of the detectors (but not the full time series or SFTs) for every segment in the observation period. Hence $\mathbf{B}'$ and $\mathbf{\Sigma}$ can be computed quickly and even faster with folded data~\cite{folding15} (in $\sim 1$ CPU minute). Thus applying our method on real data is straightforward and adds negligible amount to the computation cost.

\subsection{Estimators for a network of detectors}

We have so far derived the expressions only for a single baseline (two detectors). It is also not difficult to incorporate a network of detectors in our formalism. Let $\mathcal{I}_1$ and $\mathcal{I}_2$ denote the pair of detectors constructing the baseline $I$ and $\orf^I(f)$ is the corresponding overlap reduction function. Then the quantities, $\mathbf{C}$, $\mathbf{N}$ and $\mathbf{K}$, defined in section~\ref{subsec:derivation}, get one extra index $I$, which can then be represented as, $C_{Itf}$, $N_{Itf}$ and $K_{Itf}^{\alpha}$ respectively. The network noise covariance matrix still remains diagonal
\begin{equation}
\bm{\mathcal{N}} \ \equiv \ \mathcal{N}_{ItfI't'f'} \ := \ \langle N_{Itf}^* \, N_{I't'f'} \rangle \ = \ \delta_{II'} \delta_{tt'} \delta_{ff'} \frac{(\Delta T)^2}{4} \, P_{\mathcal{I}_1}(t;|f|) \, P_{\mathcal{I}_2}(t;|f|) \, ,
\label{eq:netNoiseCov}
\end{equation}
as two different baselines have at least one pair of different detectors with uncorrelated noise whose expectations are zero. The form of the convolution equation [Eq.~(\ref{eq:conv})] also remains the same except for the extra index $I$ (which acts in a similar way as $t$ and $f$ indices in the convolution equation). Hence the final solution takes the same form as in Eq.~(\ref{eq:defB} \& \ref{eq:normdeconv}), except that $M_{\alpha\beta}$ and $X_\alpha$ have to be redefined in the the following way,
\begin{eqnarray}
M_{\alpha\beta} &:=& \sum_{Itf} |\orf^I(|f|)|^2 \, \frac{\mathcal{F}^\alpha(|f|) \, \mathcal{F}^\beta(|f|)}{f^6 \, P_{\mathcal{I}_1}(t;|f|) P_{\mathcal{I}_2}(t;|f|)} \, , \\
Y_\alpha &:=& \left(\frac{20 \pi^2}{3 H_0^2} \right) \frac{1}{\Delta T \, M_{\alpha\alpha}} \sum_{Itf} \orf^I(|f|) \, \frac{\mathcal{F}^\alpha(|f|) \, \widetilde{s}^*_{\mathcal{I}_1}(t;f) \, \widetilde{s}_{\mathcal{I}_2}(t;f)}{|f|^3 \, P_{\mathcal{I}_1}(t;|f|) P_{\mathcal{I}_2}(t;|f|)} \, . \label{eq:networkX}
\end{eqnarray}
Thus, it is straightforward to combine single baseline estimates to get the optimal network estimate. One can simply add $M_{\alpha\beta}$ from all the baselines to get the network $M_{\alpha\beta}$ and do the same for $Y_\alpha$ with the updated network $M_{\alpha\alpha}$, rest of the steps remaining unchanged.

\section{Toy Cases}
\label{sec:toy}

We now demonstrate our method using two toy cases. For a meaningful evaluation, we inject realistic values of $\OmComp$, so that the relevance of our method in the present context can be clearly seen. In this part we restrict ourselves only to the two LIGO detectors at Hanford and Livingston. For the numerical computation we assume that the shapes of the background spectra are described by power laws, $F^\alpha = (f/f_\text{ref})^\alpha$, where $f_\text{ref} = 100$Hz is the reference frequency. The upper cut-off frequency is fixed at $512$~Hz following standard analysis, as the overlap reduction function for this detector pair accepts negligible amount of power beyond this frequency. The lower cut-off frequency is chosen as $40$~Hz for LIGO-I sensitivity and $10$~Hz for advanced detector sensitivity. Total observation time is taken as $100$ sidereal days.

In the real analysis $Y_\alpha$ are obtained by averaging over a large number of data points\footnote{To estimate $\OmComp$ from LIGO's fifth science run data the averaging was done using $\sim 1$ million segments and each segment was integrated over $\sim 2000$ independent frequency bins, so a total of $\sim 2$ billion numbers were averaged over.} [see Eq.~(\ref{eq:normX})] and $\hat{\Omega}_\alpha$ are obtained by applying a linear operator ($\mathbf{B}^{-1}$) on $Y_\alpha$, hence one can invoke the central limit theorem to argue that both $Y_\alpha$ and $\hat{\mathbf{\Omega}}$ follow multivariate Normal distributions. Since we also compute the noise covariance matrix $\mathbf{\Sigma}$ of $\hat{\mathbf{\Omega}}$, we are able to plot the $1\sigma$ (dark) and $2\sigma$ (light) contours for a pair of $\hat{\Omega}_\alpha$ (in figure~\ref{fig:twoidx} \& \ref{fig:threeidx}). For more than two indices, one can marginalise over all but a pair of indices to plot the contours in two dimensions and all but one index to get a one dimensional distribution and the $95\%$ ($2\sigma$) confidence upper limit on the corresponding $\OmComp$. All the errorbars in the plots below show $\pm 2\sigma$ bounds.

We do not add any random noise in this exercise (though we account for the noise PSDs). This is because our aim is to show that a single component search introduces bias in the estimates as well as in the errorbars (and, hence, the upper limits), which can be avoided by performing a joint analysis. If we performed hundreds of simulations by adding random Gaussian noise to $Y_\alpha$, the single and joint analysis estimates would follow the distributions which we already know, which are represented by the errorbars for single component analysis and the contours in the joint analysis, hence no extra information could be gathered through the simulations.

\subsection{Two Component Background with LIGO-I Sensitivity}
\label{subsec:twoidx}

We first consider a two component background with spectral indices $0$ and $3$. In the past, most of the searches for stochastic background using data from groundbased detectors used these two indices. In this case we use LIGO-I science requirements document (SRD)~\cite{LIGO-SRD} sensitivity. We inject the amplitudes $\Omega_0 = 5 \times 10^{-6}$ and $\Omega_3 = -5 \times 10^{-6}$ to make the upper-limits on $\Omega_0$ and $\Omega_3$ close to those obtained from LIGO's fifth science (S5) run data~\cite{sgwbS5iso}. A negative injection value of $\Omega_3$ may seem unrealistic, but our aim is to inject the unbiased estimates obtained from single index searches, where the point estimate can be negative due to statistical fluctuations\footnote{For example, in the latest LIGO-Virgo analysis the estimates for $\OmComp$ are negative for some values of $\alpha$~\cite{isoLVC_2009-10}. This may seem unphysical, but it is a normal outcome, as with the errorbars the results are consistent with zero.}. The coupling matrix equates to
\begin{equation}
\mathbf{B}' \ = \ \left[ \begin{array}{cc}
1.0000  &  0.6403 \\
0.6403  &  1.0000
\end{array} \right] \, .
\end{equation}
The error ellipses, their one dimensional marginalised distributions, mean and $95$\% confidence upper limits are shown in figure~\ref{fig:twoidx}. The relevant numbers are listed in table~\ref{tab:twoidx}.

Important point to notice in these results is that single and joint estimations do lead to different upper limits on $\Omega_{0}$ and $\Omega_{3}$. The single index upper limits on $(\Omega_0, \Omega_3)$ are $(6.901 \times 10^{-6},  3.228 \times 10^{-6})$, while those from the joint analysis are $(11.71 \times 10^{-6}, 1.61 \times 10^{-6})$. Joint estimation infers the injected value with very good accuracy---the injection point, marked with a ``$\times$'', is nearly coincident with the center of the errorbars. This is because, the only error here arises from numerical inaccuracies in the $2 \times 2$ matrix inversion.

%
\begin{figure}[h!]
    \centering
    \includegraphics[width=0.8\textwidth]{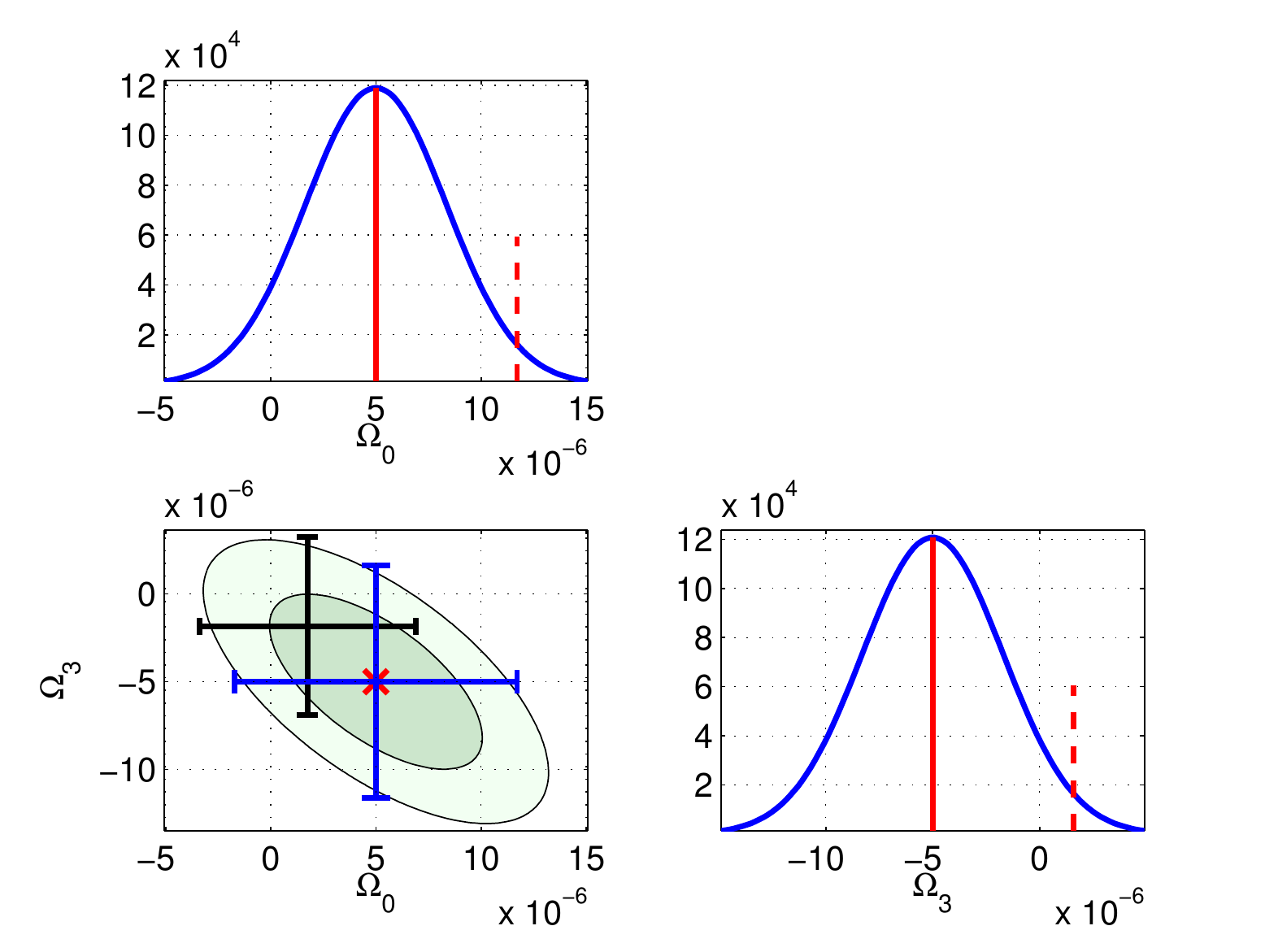}
    \caption{\label{fig:twoidx} The estimates for amplitudes of a two component background, $(\Omega_0, \Omega_3)$, are shown in this figure. The red ``$\times$'' mark shows the injected values of the parameters. The injection values are chosen in such a way that the single component analysis  upper limits are comparable to those from a recent set of upper limits from the ground based laser interferometric detectors~\cite{sgwbS5iso}. A joint analysis recovers it almost exactly (except for negligible numerical errors) by construction. The black ($2\sigma$) errorbars show the results from a single component analysis, while the blue ones show the same from the joint analysis. The contours represent the $1\sigma$ and $2\sigma$ error ellipses, corresponding to respectively $68$\% and $95$\% confidence intervals. The one dimensional plots show the marginalised probability distributions for single parameters, where the solid vertical line represents the mean, while the dashed vertical limits indicates the $2\sigma$ upper limit. The plot illustrates that single component estimates have biased mean and errorbars (and, hence, biased upper limits).}
\end{figure}
\begin{table}[h!]
\centering
\begin{tabular}{c|c|c|c|c|c|c|c}\hline\hline
 &  & \multicolumn{3}{|c|}{Single index analysis} & \multicolumn{3}{|c}{Joint multi-index analysis} \\\hline
Parameter & Injected & Estimate & $\sigma$ & $2\sigma$ upper limit & Estimate & $\sigma$ & $2\sigma$ upper limit\\\hline\hline
$\Omega_0$ & 5.0 & 1.750 & 2.575 & 6.901 & 5.0 & 3.353 & 11.71 \\\hline
$\Omega_3$ & -5.0 & -1.846 & 2.537 & 3.228 & -5.0 & 3.303 & 1.61 \\\hline\hline
\end{tabular}
\caption {\label{tab:twoidx} Summary of the injected and recovered parameters, the error bars and the $95\%$ upper limits for the two component case shown in figure~\ref{fig:twoidx}. All the numbers provided in this table are in the units of $10^{-6}$.}
\end{table}

\subsection{Three Component Background with AdvLIGO Sensitivity}
\label{subsec:threeidx}

We next consider the two LIGO detectors again, but with advanced detector sensitivity~\cite{AdvLIGO-SRD}. We inject three smaller $\OmComp$ values $(10^{-8}, 10^{-8}, 10^{-8})$ at indices $(0, 2/3, 3)$. The index $2/3$ is the predicted power-law index for a background created by compact binaries~\cite{MandicEtAl12}. The coupling matrix in this case is given by
\begin{equation}
\mathbf{B}' \ = \ \left[ \begin{array}{ccc}
 1.0000  &  0.9775   & 0.5401 \\
    0.9775  &  1.0000   & 0.6523 \\
    0.5401  &  0.6523  &  1.0000
\end{array} \right] \, .
\end{equation}
The single index search upper limits are given by $(1.68 \times 10^{-8}, 3.84 \times 10^{-8}, 1.935 \times 10^{-7})$, which are very different from those obtained from the joint analysis $(2.352 \times 10^{-8}, 4.439 \times 10^{-8}, 7.568 \times 10^{-8})$. The marginalised error ellipses, their one dimensional marginalised distributions, mean and $95$\% confidence upper limits are shown in figure~\ref{fig:threeidx} and all the relevant numbers are listed in table~\ref{tab:threeidx}.

\begin{figure}[h!]
    \centering
    \includegraphics[width=0.95\textwidth]{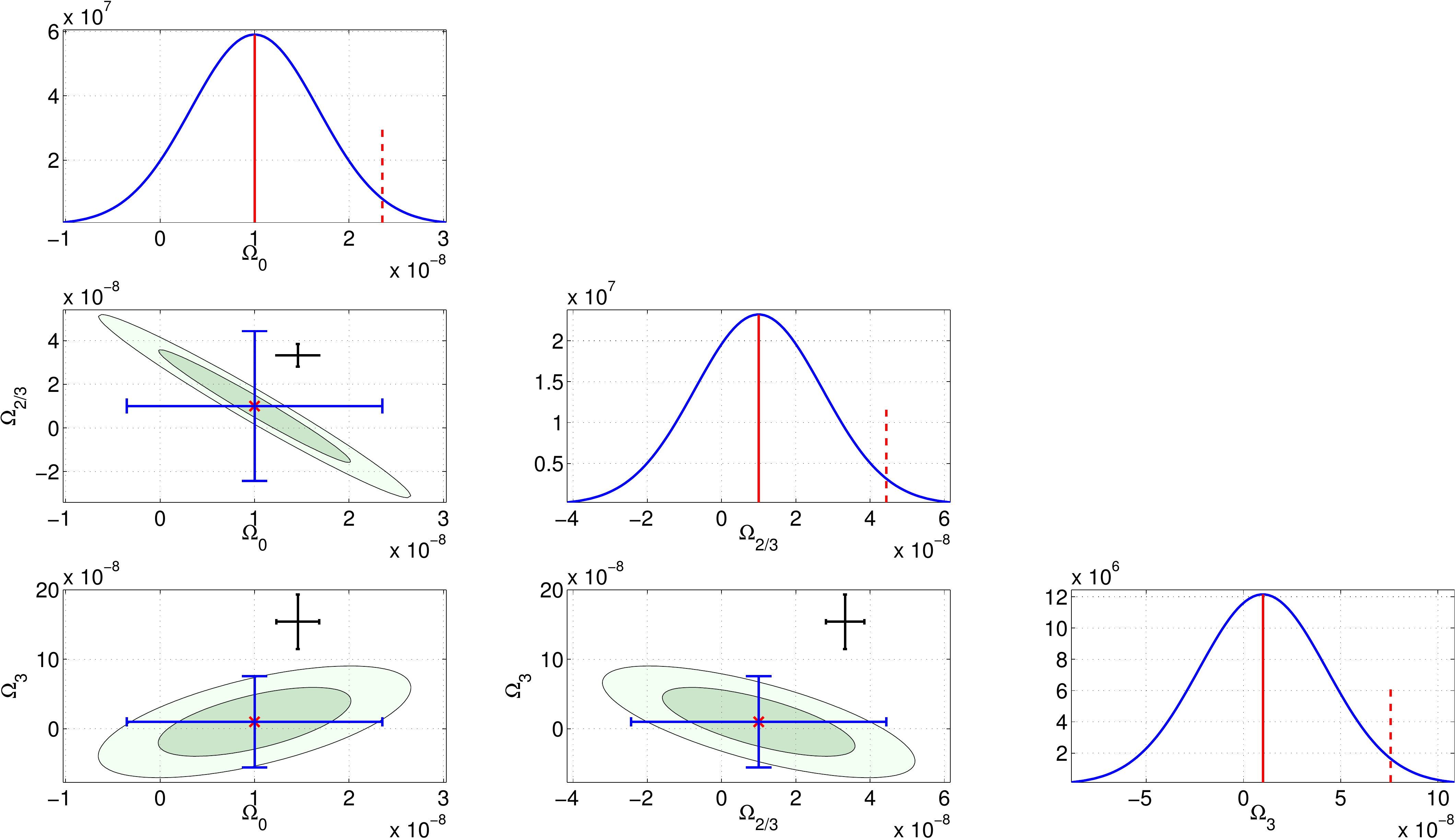}
    \caption{\label{fig:threeidx} This figure is the same as figure~\ref{fig:twoidx} expect for that here we have used an advanced LIGO design sensitivity~\cite{AdvLIGO-SRD} and the number of components is three corresponding to the spectral indices $0, 2/3, 3$. This figure shows that the estimates and the upper limits can get more severely biased in a single component search when the actual signal is close to or more than the detectable limits.}
\end{figure}
\begin{table}[h!]
\centering
\begin{tabular}{c|c|c|c|c|c|c|c}\hline\hline
 &  & \multicolumn{3}{|c|}{Single index analysis} & \multicolumn{3}{|c}{Joint multi-index analysis} \\\hline
Parameter & Injected & Estimate & $\sigma$ & $2\sigma$ upper limit & Estimate & $\sigma$ & $2\sigma$ upper limit\\\hline\hline
$\Omega_0$ & 1.0 & 1.46 & 0.113 & 1.68 & 1.0 & 0.676 & 2.352 \\\hline
$\Omega_{2/3}$ & 1.0 & 3.32 & 0.259 & 3.84 & 1.0 & 1.719 & 4.439 \\\hline
$\Omega_3$ & 1.0 & 15.41 & 1.971 & 19.35 & 1.0 & 3.284 & 7.568 \\\hline\hline
\end{tabular}
\caption {\label{tab:threeidx} Summary of the injected and recovered parameters, the error bars and the $95\%$ upper limits for the three component case shown in figure~\ref{fig:threeidx}. All the numbers provided in the table are in the units of $10^{-8}$.}
\end{table}

\section{How many components can be separated from a given dataset?}
\label{sec:indcomp}

We showed that our method can successfully estimate the amplitudes of multiple components. But how many components can be separated? If we keep on increasing the number of components to probe, the coupling matrix would likely become degenerate (two rows or columns will be nearly equal) and cannot be inverted. Our formalism provides some ground to perform a semi-quantitative study to address this question.

Since the degeneracy of the problem (coupling between estimators for different component) is adequately described by the matrix $\mathbf{B}'$ (a.k.a. the ``kernel''), we study the characteristics of this matrix. We first compute $\mathbf{B}'$ for a case with more than ten power law components, specifically, the index $\alpha$ takes values from $-1$ to $5$ with an interval of $0.5$ (total $13$ indices) which covers perhaps all the predicted cosmological and astrophysical models for the background~\cite{CowTan06}. The matrix is plotted in figure~\ref{fig:coupling}.
\begin{figure}[h!]
    \centering
    \includegraphics[width=0.8\textwidth]{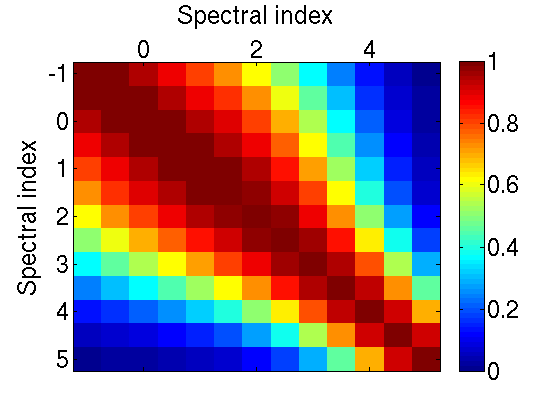} %
    \caption{\label{fig:coupling} The preconditioned coupling matrix ($\mathbf{B}'$) for the LIGO Hanford and Livingston pair with advanced detector sensitivity is plotted here. By construction the diagonals are unity, while the off digonal components are smaller positive numbers. Searching for more components makes the grid finer, keeping the overall pattern formed by the colours similar, so that, successive rows and columns of the matrix become closer and the determinant of the matrix reduces, making it ill-conditioned. The number of independent components in the matrix can be roughly estimated from singular value decomposition, shown in figure~\ref{fig:netperform}.}
\end{figure}
It can be seen that the adjacent rows or columns of the matrix are not very different, causing the matrix to become degenerate and difficult to invert. The determinant of the matrix is extremely low $\sim 10^{-60}$. This indicates that the nearby indices do not create significant difference in the signals expected in the detectors, hence they can not be distinguished with the given characteristics of the data. The problem becomes more difficult to tackle when the number of components being probed is also large, leading to progressively lower value of the determinant.

A standard technique which is often used to study the number of components which a kernel may be able to distinguish is singular value decomposition (SVD). If the ratio of a singular value to the maximum singular value is too low, the corresponding component makes it very difficult to invert the kernel, which must be discarded in order to invert the matrix~\cite{NR}. Thus, if one divides all the singular values by the maximum one, the number of ratios above a certain threshold indicates the number of degrees freedom that can be ``resolved'' with the given kernel.

We plot the singular values of $\mathbf{B}'$ in the left panel of figure~\ref{fig:netperform} for the baseline formed by the two LIGO detectors (``H'' and ``L''). To study how introduction of multiple detectors improve the ability to resolve multiple components, here we also overlay results for the networks formed by the addition of the Virgo detector (``V'') and/or a detector at an {\em arbitrary location} in India (``I'') with latitude $\sim 20^\circ$N, longitude $\sim 80^\circ$E and one arm {\em arbitrarily} oriented at $\sim 30^\circ$ West of North. The LIGO detectors and the Indian detector are assumed to have advanced LIGO sensitivity~\cite{AdvLIGO-SRD} and the Virgo detector to have advanced Virgo sensitivity~\cite{AdvVirgo-SRD}. The upper cut-off frequency was increased to $2000$~Hz, as the LIGO-Virgo pairs are known to perform much better than the LIGO-LIGO pair at higher frequencies~\cite{VirgoORF}.  In the right panel of figure~\ref{fig:netperform} we plot the expected standard deviation on the component amplitudes (ignoring their covariances) for different network configurations.
\begin{figure}[h!]
    \centering
    \includegraphics[width=0.475\textwidth]{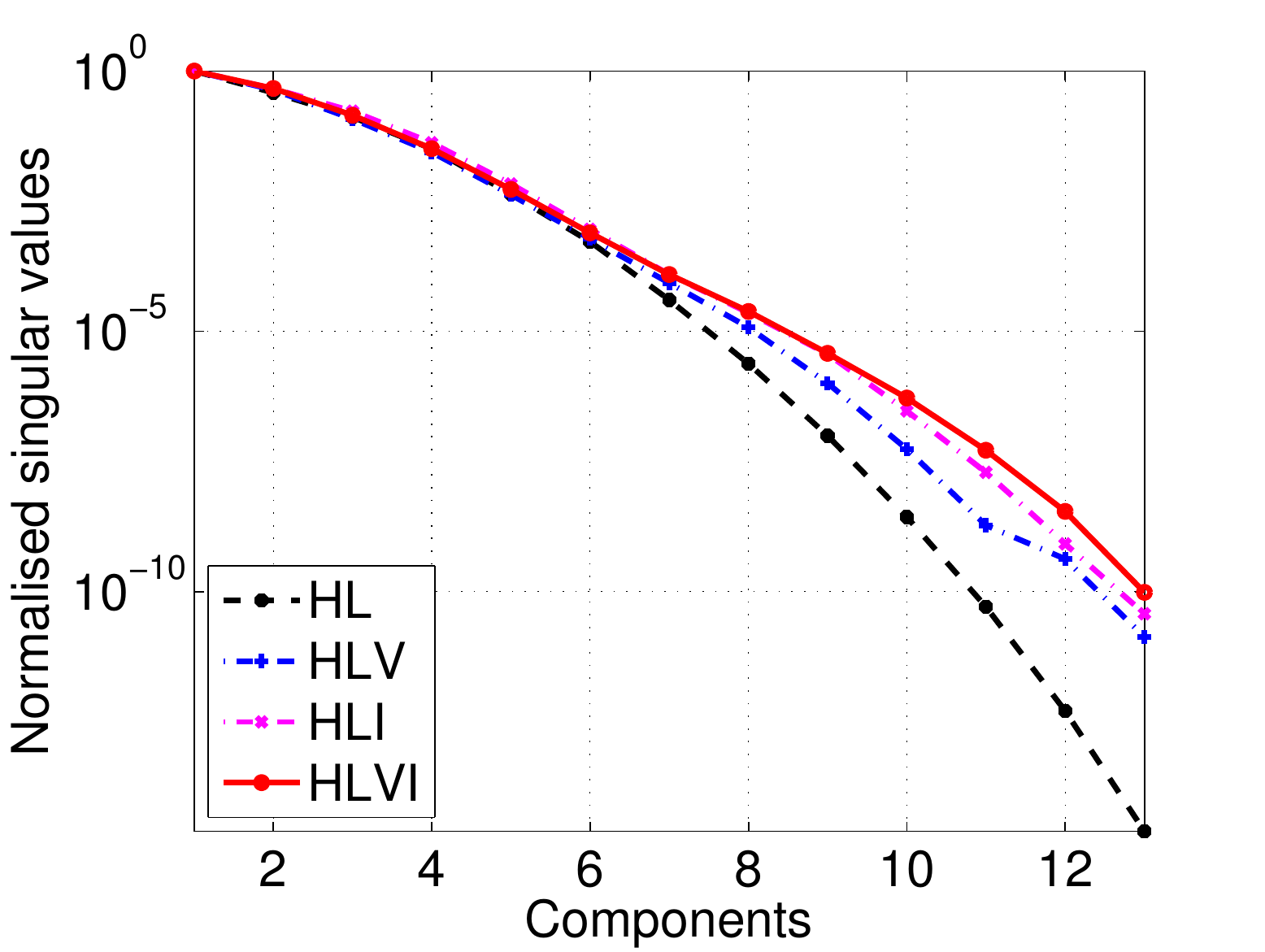} %
    \includegraphics[width=0.475\textwidth]{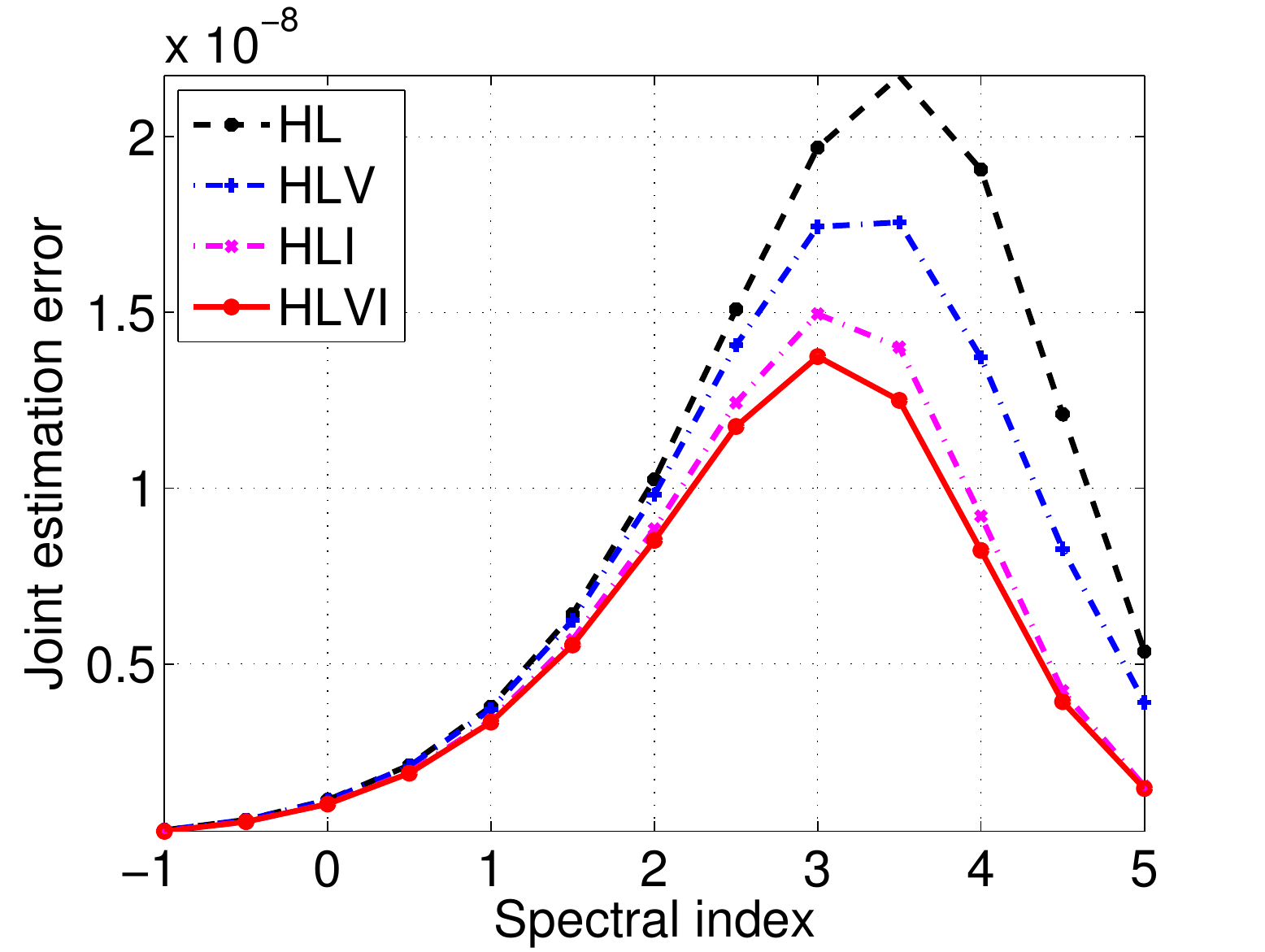} %
    \caption{\label{fig:netperform} Normalised singular values of the preconditioned coupling matrix ($\mathbf{B}'$) for different detector sets are shown on the left panel and the corresponding expected error ($\Sigma_{\alpha\alpha}^{1/2}$) on the component amplitudes (ignoring their covariances) are shown on the right panel. Slower the singular value curve falls, higher the number of components that can be estimated. Clearly, addition of detectors to the network improves the number of components that can be jointly probed and reduces the expected error on the estimators for a fixed set of components.}
\end{figure}
The plot indicates that the two LIGO detectors can jointly search for at most $10$ power-law components in this range of spectral index, while a network can increase the number by few. The expected errors also decrease with the addition of detectors in the network.

\section{Conclusions and Discussions}
\label{sec:concl}

We propose an efficient method to jointly constrain the amplitudes of multiple components of a background with known spectral shapes. Our method analytically transforms the Maximum Likelihood estimation problem to a linear deconvolution problems with dimension equal to the number of components, which is very easy to invert, as long as the data contains enough independent information on each of the components being probed. The method not only requires negligible amount of computation, but it will also be easy to apply on real data. We also investigated how many components can be jointly probed using a network of  ground based laser interferometric detectors.

Since the expected backgrounds are buried deep inside noise, in order to extract information about them from observed data, one needs to use prior information on the backgrounds. In our method we need to know the spectral shapes of the components of the background and we estimate their amplitudes. Which is the case even for the standard single index search, the index is assumed to be known. One could ask how accurately one needs to know the spectral indices. Since the expected signals in the detectors do not change much due to a change of the spectral index (by $\lesssim 1/2$), one can guess that an approximate knowledge of the spectral shape of the backgrounds would suffice to extract reasonably accurate results using our method. Also, it is trivial to incorporate arbitrary non-power-law spectral profiles in this scheme, even in the numerical implementation. We have used power law spectra here solely because it is the usual choice in the current analyses, even though theoretical models not necessarily predict power laws.

Increasing the number of components to probe increases the error on each of the components, as the data can only offer a finite amount of information. Because of this, a single component search offers the smallest errorbar for a given spectral shape. One can, however, combine the estimates for multiple backgrounds optimally, to extract higher signal to noise ratios for detection purposes. This can also involve inclusion of relative weights motivated by a certain cosmological scenario.
One may be able to take one step further and perform an analysis to obtain posterior distributions of the amplitudes for the given prior.

Fine tuning an estimation process can be a never ending endeavour. For immediate application on data from the advanced ground based interferometers, targeting perhaps three or four components would be a good starting point. Since higher the number of probed components more the error, one can increase the number as data quality improves and more detectors in the network come online. In the end, making such choices should become easier with experimentation on real data.

\acknowledgments{
We would like to thank the stochastic analysis group of the LIGO-Virgo Collaboration for useful discussions. SM acknowledges the support of the Science and Engineering Research Board (SERB), India through the fast track grant SR/FTP/PS-030/2012. SJ acknowledges support from grant under ISRO-RESPOND program (ISRO/RES/2/384/2014-15).}

\bibliographystyle{JHEP}
\bibliography{multibgiso}

\end{document}